\documentclass{JHEP3}
\usepackage{amsmath}
\usepackage{cite}
\usepackage{epsfig}

\newcommand{\cL} {{\cal L}}
\newcommand{\BB} {{\cal B}}
\newcommand{\cO} {{\cal O}}
\newcommand{\al} {{\alpha}}
\newcommand{\la} {{\lambda}}
\newcommand{\La} {{\Lambda}}

\newcommand{\ket}[1] {\left|#1\right>}

\title{Marginal deformation for the photon in superstring field theory}
\author{Ehud Fuchs, Michael Kroyter\\
Max-Planck-Institut f\"ur Gravitationsphysik\\
Albert-Einstein-Institut\\
14476 Golm, Germany\\
\email{udif@aei.mpg.de, mikroyt@aei.mpg.de}
}

\abstract{We find solutions of supersymmetric string
field theory that correspond to the photon marginal deformation
in the boundary conformal field theory.
We revisit the bosonic string marginal deformation and generate
a real solution for it. We find a map between the solutions of
bosonic and supersymmetric string field theories and
suggest a universal solution to superstring field theory.
}

\preprint{AEI-2007-042}

\begin{document}

%=====================
\section{Introduction}
%=====================

Our understanding of open bosonic string field theory~\cite{Witten:1986cc}
has deepened following Schnabl's analytic solution~\cite{Schnabl:2005gv},
as can be seen from the papers that
followed~\cite{Okawa:2006vm,Fuchs:2006hw,Fuchs:2006an,Rastelli:2006ap,
Ellwood:2006ba,Fuji:2006me,Fuchs:2006gs,Okawa:2006sn,Imbimbo:2006tz,
Erler:2006hw,Erler:2006ww,Kishimoto:2007bb}.
Specifically, marginal deformations~\cite{Sen:1990hh,Sen:1990na,Sen:1992pw}
were found in~\cite{Kiermaier:2007ba,Schnabl:2007az} and used
to study the rolling tachyon in~\cite{Ellwood:2007xr}
(for earlier related works
see~\cite{Zwiebach:2000dk,Iqbal:2000qg,Takahashi:2002ez,Kluson:2002hr,
Katsumata:2004cc,Sen:2004cq,Yang:2005iu,Kishimoto:2005bs}).
A different approach to generate such solutions was given
in~\cite{Fuchs:2007yy}.
The added value of the new approach is that the
handling of marginal deformations that correspond to operators with
a singular OPE is much easier.

For open superstring field
theory~\cite{Berkovits:1995ab,Berkovits:2000hf,Berkovits:2001nr}
marginal deformation solutions 
were found in~\cite{Erler:2007rh,Okawa:2007ri,Okawa:2007it}.
The purpose of this paper is to generalize the methods of~\cite{Fuchs:2007yy}
to the supersymmetric theory. This gives the marginal deformation solution
corresponding to the photon whose OPE is singular.

There is a similarity between pure gauge solutions to the bosonic
Chern-Simon-like theory and the equation of motion of the supersymmetric
WZW-like theory.
This similarity was recently used to generate superstring
marginal deformations~\cite{Erler:2007rh}.
To adapt the method of~\cite{Fuchs:2007yy} from the bosonic string
to the superstring, we exploit a different similarity, one which relates the
pure gauge solutions of the two theories.

The results in~\cite{Fuchs:2007yy} rely on the fact that solutions of
bosonic string field theory can be formally written as pure gauge solutions
\begin{equation}
\label{PsiForm}
\Psi = \Gamma(\Phi)^{-1}Q\Gamma(\Phi)\,,
\end{equation}
where $Q$ is the BRST charge.
In the relation above, $\Phi$ is a string field from which the
solution is generated and $\Gamma(\Phi)$ is a function of the form
\begin{equation}
\label{fForm}
\Gamma(\Phi)=1+\Phi+\cO(\Phi^2)\,,
\end{equation}
where the product used is the star product and the ``1'' represents
the identity state, which guarantees that the function $\Gamma(\Phi)$
is invertible.
The first order term generates the linear gauge transformation $Q\Phi$.

The reason that $\Psi$ is a physical solution
has to do with the singular nature of $\Phi$.
In~\cite{Fuchs:2007yy}, the singularity for the photon marginal deformation
was due to the linear dependence of $\Phi$
on $x_0$, which is the zero mode of the scalar field $X$
(space-time indices are not explicitly mentioned throughout the paper
as we always work with a single space-like scalar field).
In order for the solution to make sense we
require that $\Psi$ is $x_0$-independent
\begin{equation}
\partial_{x_0}\Psi = \partial_{x_0}\big(\Gamma(\Phi)^{-1}Q\Gamma(\Phi)\big)=0\,.
\end{equation}
This resembles the equation of motion of the superstring
field~\cite{Berkovits:1995ab,Berkovits:2000hf}
\begin{equation}
\label{SusyEom}
\eta_0\big(G^{-1}QG)=0\,,
\end{equation}
where $\eta_0$ is the superstring ghost field zero mode,
which behaves as a second BRST charge
(more details on the superstring field theory we are using are
given in section~\ref{sec:SusyIntro}).
However, this similarity is not the one we wish to exploit.
Instead, we wish to compare the pure gauge solutions of both theories.
In the supersymmetric theory
the infinitesimal gauge transformation
depends on two gauge fields $\Lambda,\tilde\Lambda$
\begin{equation}
\label{InfiGauge}
\delta G = -(Q\tilde\Lambda)G + G(\eta_0\Lambda)\,.
\end{equation}
The integrated form of this infinitesimal gauge transformation is
\begin{equation}
G_\lambda = e^{-\lambda Q\tilde\Lambda} G_0 e^{\lambda\eta_o\Lambda}\,.
\end{equation}
The two exponents above may be replaced by any functions  of the
form~(\ref{fForm}), since this corresponds to a field redefinition.
Because we are interested in pure gauge solutions, $G_0$
is the identity state and $G$ is
\begin{equation}
\label{GForm}
G=\tilde\Gamma(\tilde\Phi)^{-1}\Gamma(\Phi)\,,\qquad
\tilde\Phi\equiv Q\tilde\Lambda\,,\quad
\Phi\equiv \eta_0\Lambda\,.
\end{equation}

The gauge field $\Phi$ of the bosonic string obviously does not have
insertions of the superstring ghost field $\xi$,
which is the conjugate of the $\eta$ ghost.
Therefore, it is $\eta_0$-exact and can be used to
define $\Lambda$ such that $\Phi$ in both theories are the same
\begin{equation}
\label{SUSYbos}
\Lambda = \xi(z)\Phi_\text{bosonic} \quad\Rightarrow\quad
\Phi_\text{SUSY} = \Phi_\text{bosonic} \equiv \Phi\,,
\end{equation}
for any $z$.
$\tilde\Phi$ can also be based on $\Phi$ in a similar way.
Here, we get a new state since $\Phi$ is not $Q$-exact
\begin{equation}
\label{Map}
\tilde\Lambda = P(z)\Phi \quad\Rightarrow\quad
\tilde\Phi = \Phi - P(z)Q\Phi\,,
\end{equation}
where $P(z)$ is the inverse operator of $Q$ described in
section~\ref{sec:SusyIntro}.
This gives a mapping of bosonic solutions to supersymmetric
ones. One gets
\begin{equation}
G=1+P\Psi+...\,,
\end{equation}
where the ellipsis stand for corrections, such as higher order
$\la$ corrections in the marginal deformation case. This mapping
seems to be a natural one, since up to higher order corrections
the ghost number zero superstring field is obtained from
the ghost number one bosonic string field by the action of the
$P$ operator.
For the marginal deformation this canonical choice does not work
because of the additional constrain of $x_0$-independence.

A short calculation shows that the bosonic solution can be written as
\begin{equation}
\label{PsiG}
\Psi(G) = G^{-1}QG\,.
\end{equation}
Therefore
it is clear that if the supersymmetric solution is $x_0$-independent
then the bosonic solution follows suit
\begin{equation}
\label{OtherDir}
\partial_{x_0}G=0 \quad\Rightarrow\quad \partial_{x_0}\Psi=0\,.
\end{equation}
The relation~(\ref{PsiG}) between solutions of the bosonic and supersymmetric
theories, is not one-to-one, different $G$'s can result in the same $\Psi$.
Thus,~(\ref{OtherDir}) does not hold in the other direction. The
$\mbox{$x_0$-independence}$ of $\Psi$ does not impose any condition on how
$\tilde\Phi$ enters $G$. However, as we will see, for any $\Psi$
one can find a corresponding $G$ that is $\mbox{$x_0$-independent}$.
We find a map $G(\Psi)$ satisfying $\partial_{x_0}G(\Psi)=0$,
such that~(\ref{PsiG}) is obeyed, that is $\Psi(G(\Psi))=\Psi$.

The rest of the paper is organized as follows.
We end this introduction by presenting the superstring field
theory that we use.
Then, in section~\ref{sec:Bosonic} we summaries the bosonic
marginal deformation solutions~\cite{Fuchs:2007yy} and extend the formalism
to generate solutions satisfying the reality condition
in~\ref{sec:realCond}.
Next, in section~\ref{sec:Map}, we find the condition for $\mbox{$x_0$-independence}$
for solutions of the supersymmetric theory.
In section~\ref{sec:Susy} we present the marginal deformation of
the superstring. First, we show in~\ref{sec:FirstOrder} that the
photon of the supersymmetric theory
can also be written as an exact state. Then, we generate a solution
to all orders in~\ref{sec:AllOrders} and a real solution in~\ref{sec:Real}.
In section~\ref{sec:Universal} we suggest that our method can
be used to generate the universal superstring solution corresponding
to Schnabl's solution~\cite{Schnabl:2005gv}. In this case it is possible
to use~(\ref{Map}).
We wrap things up with conclusions in~\ref{sec:Conclusions}.
In the appendices we relate our solutions to those
of~\cite{Erler:2007rh,Okawa:2007ri}.

%====================================
\subsection{Superstring field theory}
%====================================
\label{sec:SusyIntro}

There are two main versions of supersymmetric open string field theory.
It was shown that the one introduced in~\cite{Witten:1986qs} suffers from
singularities due to collision of picture-changing operators~\cite{Wendt:1987zh}.
While it is plausible that a modification of this theory of the form presented
in~\cite{Preitschopf:1989fc,Arefeva:1989cm,Arefeva:1989cp} can save the day, we
prefer to use the more established superstring field theory, due to
Berkovits~\cite{Berkovits:1995ab,Berkovits:2000hf,Berkovits:2001nr}.

The action in the NS sector is a generalization of WZW theory, where the two
differentials $\partial, \bar \partial$ are replaced by $Q,\eta_0$. The action
can be extended to include also the Ramond sector, albeit not in a
Lorentz-covariant way.
Only fields in the NS sector get a vev.
The equation of motion for the NS sector is~(\ref{SusyEom}).
The string field $G$ in this equation depends on the variables $X^\mu,\psi^\mu$,
as well as on the $b,c$ ghosts and on the ``bosonized''
superghosts~\cite{Friedan:1985ge}.
The bosonization is given by
\begin{equation}
\beta=\partial\xi e^{-\phi}\,,\qquad\gamma=\eta e^{\phi}\,,
\end{equation}
where $\xi,\eta$ are a conjugate pair of fermions and $\phi$ is a scalar field,
such that the solitons $e^{\pm\phi}$ are fermions.
It is clear that the superghosts do not depend on the
zero mode of $\xi$. Thus, the physical space is in the ``small Hilbert space'',
not including the zero mode, while the ``large Hilbert space'' contains one more
copy of the small one with $\xi_0$ acting on it.

An important peculiarity of the RNS string is the existence of an infinity
of vertex operators for any given physical state. The various ``pictures''
of the vertex operators are easy to understand using the $\xi,\eta,\phi$ variables.
Each operator is assigned a picture-number as in table~\ref{table}.
\TABLE{
\label{table}
\begin{tabular}{c|crr}
operator & $h$ & $n_g$ & $n_p$ \\
\hline
$b$ & 2 & -1 & 0 \\
$c$ & -1 & 1 & 0 \\
$\eta$ & 1 & 1 & -1 \\
$\xi$ & 0 & -1 & 1 \\
$e^{q\phi}$ & $-\frac{q(q+2)}{2}$ & 0 & $q$ \\
$\beta=\partial\xi e^{-\phi}$ & $\frac{3}{2}$ & -1 & 0 \\
$\gamma=\eta e^{\phi}$ & $-\frac{1}{2}$ & 1 & 0 \\
$J_B$ & 1 & 1 & 0 \\
$P$ & 0 & -1 & 0 \\
\end{tabular}
\caption{The conformal weight $h$, ghost number $n_g$ and picture number
$n_p$ of the superstring field theory operators we work with.}
}

Now, given
a vertex operator $V$, another vertex operator describing the same state, but in
a different picture, is obtained by~\cite{Friedan:1985ge}
\begin{eqnarray}
	\tilde V=[Q,\xi V]\,.
\end{eqnarray}
While this state is exact in the large Hilbert space, it is only closed in the
small Hilbert space, due to the appearance of $\xi_0$ in its definition.
Scattering amplitudes can be calculated with
any set of representatives from the equivalence classes of vertex operators, as
long as the picture number is exactly saturated, as shown in~\cite{Friedan:1985ge}.

In Berkovits' string field theory, the states live in the large hilbert space.
However, since $\eta_0$ acts as a generator of gauge transformation (in the linearized
theory), the whole small hilbert space is composed of gauge degrees of freedom.
Since the other part of the space is just $\xi_0$ times the small hilbert space,
it has exactly the correct amount of degrees of freedom to represent the string
with $Q$ as the BRST operator. In this way the theory is described without the need
to use explicit picture changing operators. Thus avoiding the potential problems of
other formulations.

An important property of the large Hilbert space is that the BRST charge
\begin{equation}
Q = \oint dz J_B(z) =
    \oint dz \Big(c(T_m+T_{\xi\eta}+T_\phi) + c\partial cb + \eta e^{\phi}G_m
    -\eta\partial\eta e^{2\phi}b\Big),
\end{equation}
has an inverse in this space
\begin{equation}
\{Q,P(z)\}=1\,,\qquad P(z)\equiv -\xi\partial\xi e^{-2\phi}c(z)\,.
\end{equation}
To verify this we use the following identities
\begin{align}
&T_{\xi\eta} = -\eta\partial\xi\,,\qquad\qquad\qquad\qquad\ \qquad\qquad
\eta(z)\xi(0) \sim \frac{1}{z}\,,\\&
\begin{aligned}
&T_\phi = -\frac{1}{2}\partial\phi\partial\phi - \partial^2\phi\,,\qquad\qquad\qquad\,\qquad
\phi(z)\phi(0) \sim -\log z\,,\\
&\phi(z) e^{q\phi(0)} \sim -q\log z e^{q\phi(0)}\,,\qquad\qquad\qquad
e^{q_1\phi(z)}e^{q_2\phi(0)} = z^{-q_1 q_2} e^{q_1\phi(z)+q_2\phi(0)}\,.
\end{aligned}
\end{align}

Some other useful identities include
\begin{equation}
Q^2 = P^2(z) = \eta_0^2 = \xi^2(z) = \{Q,\eta_0\} = 0\,,\qquad
\{\eta_0,\xi(z)\} = 1\,.
\end{equation}
These relations reveal a duality under exchange of $Q$ with $\eta_0$
and $P(z)$ with $\xi(z)$.

%======================================
\section{Revisiting the bosonic string}
%======================================
\label{sec:Bosonic}

%=======================================
\subsection{Photon marginal deformation}
%=======================================

The bosonic string marginal solution of~\cite{Fuchs:2007yy} was based on
the fact that the physical photon state can be written as an exact state
\begin{equation}
\Psi_1=c\partial X(0)\ket{0}=QX(0)\ket{0}.
\end{equation}
This means that any pure gauge string field~(\ref{PsiForm}),
which automatically satisfies the equation of motion, is a candidate
photon marginal solution provided that it generates the
first order state. This only requires $\Gamma(\Phi)$ to be of the
form~(\ref{fForm}) and
\begin{equation}
\Phi = \lambda X(0)\ket{0}+\cO(\lambda^2).
\end{equation}
We refer to different choices of $\Gamma(\Phi)$ as
``different schemes''~\cite{Fuchs:2007yy}.
The solution to linear order is scheme independent,
and we can generate identical solutions to all orders using different
schemes by modifying the higher order terms of $\Phi$.

For a solution to be meaningful, it also has to be $x_0$-independent.
This can be achieved by an appropriate choice of the non-linear terms
of $\Phi$. We refer to such terms as counter terms.
For the ``left'' and ``right'' schemes
\begin{align}
\Gamma_L(\Phi)=\frac{1}{1-\Phi}\quad\Rightarrow\quad
\Psi_L = (1-\Phi)Q\frac{1}{1-\Phi}\,,\\
\Gamma_R(\Phi)=1+\Phi\quad\Rightarrow\quad
\Psi_R = \frac{1}{1+\Phi}Q(1+\Phi)\,,
\end{align}
we have an explicit expression for the counter terms that generates
such a solution.
Concentrating on $\Psi_L$, it would be $x_0$-independent, provided
that $\Phi$ satisfies the linear differential equation
\begin{equation}
\label{LeftDE}
\partial_{x_0}\Phi = \lambda(1-\Phi)\Omega\,,
\end{equation}
where $\Omega$ is the vacuum state.

It is the specific form of the function $\Gamma_L(\Phi)$,
which allows us to easily calculate derivatives despite the
fact that we are working with a
non-commutative algebra
\begin{equation}
\partial \frac{1}{1-\Phi} = \frac{1}{1-\Phi} \partial\Phi \frac{1}{1-\Phi}\,,
\end{equation}
where $\partial$ can stand for any derivation.
This gives
\begin{align}
\partial_{x_0}\Psi_L &= -\partial_{x_0}\Phi Q\frac{1}{1-\Phi}
    + (1-\Phi)Q\Big(\frac{1}{1-\Phi}\partial_{x_0}\Phi\frac{1}{1-\Phi}\Big)
\nonumber\\
    & = -\lambda(1-\Phi)\Omega Q\frac{1}{1-\Phi}
    + \lambda(1-\Phi)Q\Big(\Omega\frac{1}{1-\Phi}\Big) = 0\,.
\end{align}
To solve~(\ref{LeftDE}) we expand
\begin{equation}
\Phi = \sum_{n=1}^{\infty}\lambda^n\Phi_n\,.
\end{equation}
This reveals that the differential equation~(\ref{LeftDE}) is actually
an infinite set of differential equations
\begin{equation}
\partial_{x_0}\Phi_1 = \Omega\,,\quad
\partial_{x_0}\Phi_n = - \Phi_{n-1}\Omega\,.
\end{equation}
Solving these linear equations order by order is straightforward,
but although there are many possible solutions, we only have one solution
in closed form
\begin{equation}
\label{leftSchemeSol}
\Phi_n = -\frac{(-1)^n}{n!}(X^n,\underbrace{1,\ldots,1}_{n-1})\,.
\end{equation}
Here we are using the $n$-vector notation to represent the wedge
state
$\ket{n+1}$~\cite{Rastelli:2000iu,Rastelli:2001vb,Schnabl:2002gg},
where the vector elements represent
the operator insertions at the $n$ canonical sites of the wedge state.
Normal ordering at each site is implicit and $1$ stands for the
identity insertion, i.e. no insertion. This is illustrated in
figure~\ref{fig:Wedge}.
\FIGURE{
\label{fig:Wedge}
\begin{picture}(0,0)%
\includegraphics{Wedge.pstex}%
\end{picture}%
\setlength{\unitlength}{3947sp}%
\begingroup\makeatletter\ifx\SetFigFont\undefined%
\gdef\SetFigFont#1#2#3#4#5{%
  \reset@font\fontsize{#1}{#2pt}%
  \fontfamily{#3}\fontseries{#4}\fontshape{#5}%
  \selectfont}%
\fi\endgroup%
\begin{picture}(7149,2700)(278,-2449)
\put(1694,-2386){\makebox(0,0)[lb]{\smash{{\SetFigFont{12}{14.4}{\familydefault}{\mddefault}{\updefault}$X^n$}}}}
\put(2968,-2386){\makebox(0,0)[lb]{\smash{{\SetFigFont{12}{14.4}{\familydefault}{\mddefault}{\updefault}1}}}}
\put(278,-2374){\makebox(0,0)[lb]{\smash{{\SetFigFont{12}{14.4}{\familydefault}{\mddefault}{\updefault}$-\frac{(n+1)\pi}{4}$}}}}
\put(5710,-2387){\makebox(0,0)[lb]{\smash{{\SetFigFont{12}{14.4}{\familydefault}{\mddefault}{\updefault}1}}}}
\put(6856,-2386){\makebox(0,0)[lb]{\smash{{\SetFigFont{12}{14.4}{\familydefault}{\mddefault}{\updefault}$\frac{(n+1)\pi}{4}$}}}}
\put(1735,-1606){\makebox(0,0)[lb]{\smash{{\SetFigFont{12}{14.4}{\familydefault}{\mddefault}{\updefault}$\frac{\pi}{2}$}}}}
\end{picture}%

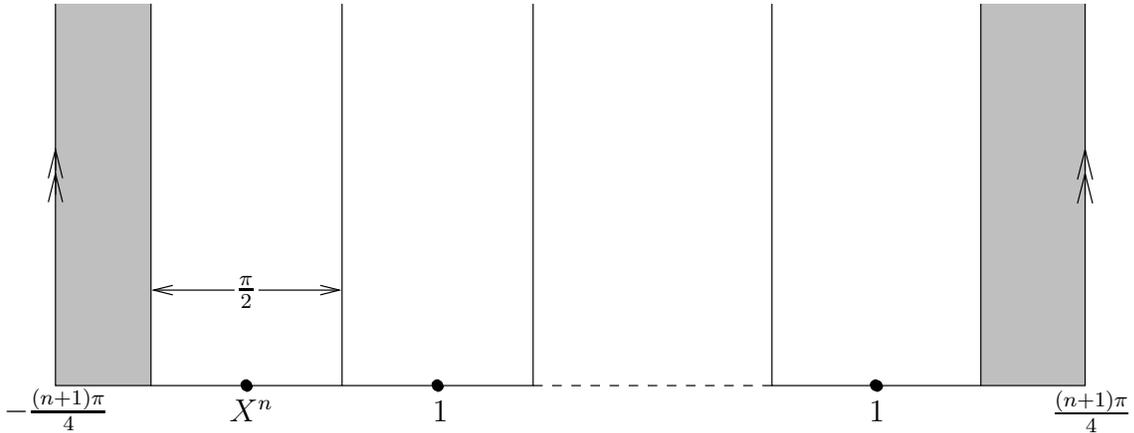
\caption{Graphical representation of the state $\Phi_n$~(\ref{leftSchemeSol}).
The worldsheet is a semi-infinite cylinder
(the double-arrowed lines are identified with each other) of circumference
$\frac{(n+1)\pi}{2}$, where the coordinate patch is marked in gray.
The canonical upper-half-plane coordinate $\xi$ is mapped to
this cylinder using the transformation
$z=\frac{n+1}{2}\arctan(\frac{2}{n+1}\xi)$.
The operator $X^n$ is a product of $n$ scalar fields $X(z)$ in the
cylinder coordinates, where normal ordering is implicit.
The 1's stand for no operator insertion and in this sense they are
redundant.
They are only presented to clarify the relation to the $n$-vector
notation in~(\ref{leftSchemeSol}).
}
}

Actually, at each order the number of degrees of freedom for generating
a solution is
\begin{equation}
\label{DOFansatz}
\dim(\Phi_n)=\binom{2n-2}{n}\,.
\end{equation}
These degrees of freedom are in general complex.
They correspond to the number of gauge degrees of freedom within our ansatz.

%=================================
\subsection{The reality condition}
%=================================
\label{sec:realCond}

Next we would like to find a solution that satisfies the string field
reality condition.
The reality condition states that hermitian conjugation and BPZ
conjugation agree~\cite{Witten:1986cc}.
In our vector notation the reality condition translates
to the following statement.
Write the state in the opposite orientation, with a factor of $(-1)$
for every $\partial X$ or $\partial c$ insertion and no factors for $X$
and $c$ and complex conjugate the
coefficients.
If this
procedure returns the original state then the state is real.
For simplicity, we consider only real coefficients,
as it turns out that this is sufficient for constructing a real string field.
In particular we choose the function $\Gamma(\Phi)$ of~(\ref{PsiForm})
to be a real function.
As we will see the reality of $\Psi$ implies that $\Phi$ should be imaginary
and this implies that {\it the deformation parameter $\lambda$
should be imaginary}.

Since $\Phi$ is built only from $X$ insertions
the reality of $\Phi$ is simply related to its symmetry.
The component $\la^n\Phi_n$
is imaginary provided it is symmetric under inversion when $n$ is odd
and antisymmetric when $n$ is even. To evaluate the number of degrees
of freedom we consider the space of solutions of the homogeneous equation
\begin{equation}
\partial_{x_0}\Phi_n=0\,.
\end{equation}
The space of solutions of this equation is given by the quotient of
the space of homogeneous polynomials of degree $n$ in $n$ variables by
the space of homogeneous polynomials of degree $n-1$ in $n$ variables.
The dimension of this space is given by~(\ref{DOFansatz}).
We now divide both spaces into symmetric and antisymmetric parts.
The derivative $\partial_{x_0}$ does not change the symmetry property
and, as we shall soon demonstrate, it is also possible to define integration
in a way that respects the symmetry. Hence, the number of (anti-)symmetric
degrees of freedom is just the dimension of the quotient space of the
two (anti-)symmetric spaces. The combinatorics is different for the cases
of $n$ odd/even. The result can be summarized by
\begin{equation}
\dim(\Phi_n^{S,A})=\frac{1}{2}\Bigg(\binom{2n-2}{n}\pm
    \frac{1+(-1)^n}{2}
      \binom{n-1}{\frac{n}{2}}\Bigg)\,,
\end{equation}
where the plus sign stands for the symmetric case.

We have only two solutions in a closed form, the one described above in the
left-scheme and a corresponding solution in the right-scheme.
It is easy to see that these solutions are not real.
We can, however, generate different solutions. Let us work in
the left-scheme.
At level two, imposing the reality condition, and
using only real coefficients the unique solution is
\begin{equation}
\Phi_2=-\frac{1}{4}\big((X^2,1)+2(X,X)-(1,X^2)\big)\,.
\end{equation}
At level three there are two degrees of freedom
for choosing a real solution.

Already in the expression for the counter terms at level two we have
the term $(X,X)$, which we interpret as
`changing the scheme'~\cite{Fuchs:2007yy}. Thus,
it may seem beneficial to start with a scheme where the symmetry is
more transparent.
We want a systematic procedure for generating real solutions.
We define $\Phi^*$ to be the string field obtained from $\Phi$ by a
combination of hermitian and BPZ conjugations.
From
\begin{equation}
\Psi^*=\Gamma(\Phi^*)Q \Gamma(\Phi^*)^{-1}\,,
\end{equation}
we see that the reality condition can be written as\footnote{
Expanding $\Gamma(\Phi)$ in $\lambda$, this condition fixes the real
part of the $n^\text{th}$ order in term of the lower orders.}
\begin{equation}
\Gamma(\Phi)=\Gamma(\Phi^*)^{-1}\,.
\end{equation}
This condition is generically non-linear in $\Phi,\Phi^*$.
However for schemes of the form
\begin{equation}
\label{RealCond}
\Gamma(\Phi)=\Gamma(-\Phi)^{-1}\,,
\end{equation}
we get that the reality condition on $\Psi$ gives the linear condition
\begin{align}
\label{imgainaryPhi}
\Phi^*=-\Phi\,.
\end{align}
It is indeed natural to require that $\Phi$ is imaginary
since $\la\Phi_1$ is imaginary.
The three other schemes that were specifically
considered in~\cite{Fuchs:2007yy}, i.e., the symmetric scheme,
the exponent scheme and the square root scheme, are given by
\begin{align}
\Gamma_S(\Phi)=\frac{1+\frac{\Phi}{2}}{1-\frac{\Phi}{2}} \,,\qquad
\Gamma_E(\Phi)=e^\Phi \,,\qquad
\Gamma_R(\Phi)=\sqrt{\frac{1+\Phi}{1-\Phi}}\,.
\end{align}
They all obey~(\ref{imgainaryPhi}).

For the symmetric scheme we can use the algebraic relation
between the two $\Phi$'s to obtain a differential equation
analogous to~(\ref{LeftDE}),
\begin{align}
\label{inductionS}
\partial_{x_0}\Phi=\la
    (1-\frac{\Phi}{2})\Omega(1+\frac{\Phi}{2})\,.
\end{align}
Note that this equation is invariant under
conjugation~(\ref{imgainaryPhi}), since the conjugate
of $\partial_{x_0}$ is $-\partial_{x_0}$.
This yields a recursion relation for $\Phi_k$,
\begin{align}
\label{inductionSco}
\partial_{x_0}\Phi_k = \frac{1}{2}\Omega\Phi_{k-1}
    - \frac{1}{2}\Phi_{k-1}\Omega
    - \frac{1}{4}\sum_{j=1}^{k-2}\Phi_{j}\Omega\Phi_{k-1-j}\,.
\end{align}

We prove that real solutions to the above equation
exist within our ansatz by providing an explicit integration
recipe that is manifestly imaginary. This not
only proves that a real solution exists, but also gives an easy
algorithm to find it order by order. In fact, one can define explicitly
infinitely many different recursion relations leading to real solutions.
We provide explicit results for the first few coefficients
one gets using some of these algorithms. We also give
a closed form expression for one of the possible recursion relations.

Given a site $k$, one can define the integration ``localized at this site''
of a length-$n$ vector by
\begin{align}
\nonumber
& \int_k (X^{j_1},..,X^{j_k},..,X^{j_n})\equiv
  \frac{1}{j_k+1} (X^{j_1},..,X^{j_k+1},..,X^{j_n})\\&-
  \frac{1}{(j_k+1)(j_k+2)} \sum_{m\neq k}\partial_{X_m}
     (X^{j_1},..,X^{j_k+2},..,X^{j_n})\\&+
\nonumber
  \frac{1}{(j_k+1)(j_k+2)(j_k+3)}
    \sum_{m_1\neq k}\partial_{X_{m_1}}\sum_{m_2\neq k}\partial_{X_{m_2}}
     (X^{j_1},..,X^{j_k+2},..,X^{j_n})-\ldots
\end{align}
The number of terms is finite, since the total power is finite.
The result of applying more than $\sum_{m\neq k}j_m$ derivatives is zero.
We are performing an integration by parts, such that the power
at the $k^{\text{th}}$ site is raised, while the power at other sites
is reduced.

Since the power of $X$ is always raised by one in the integration,
the combination $\la(\int_k+\int_{n-k})$ is imaginary
(recall that the number of $X$'s equals the number of
the $\la$'s in $\Phi$ and that $\la$ is imaginary).
The integration operations
are linear and so {\it any} combination of the form
\begin{align}
\int_{\vec\al_n}\equiv\sum_{k=1}^n \al_n^k\int_k \,,\qquad \qquad
     \sum_{k=1}^n \al_n^k=1\,,\qquad \qquad a_n^k=a_n^{n+1-k}\,,
\end{align}
yields an imaginary integration prescription, which
leads to a well defined recursion relation.
Integrating~(\ref{inductionSco}) using such a recursion relation
gives a solution that is imaginary by construction.

For example, the choice
$\al_n^1=\al_n^n=\frac{1}{2},\ \al_n^{k\notin \{1,n\}}=0$,
gives at the first few orders
\begin{align}
\Phi_1 =& (X)\,,\\
\Phi_2 =& \frac{1}{4}\Big((1,X^2)-(X^2,1)\Big)\,,\\
\Phi_3 =& \frac{1}{48}\Big((X^3,1,1)+6(X^2,X,1)-3(X^2,1,X)-6(X,X^2,1)
\nonumber\\
    &\qquad -3(X,1,X^2)-6(1,X^2,X)+6(1,X,X^2)+(1,1,X^3) \Big).
\end{align}
Another simple choice is
$\al_n^k=\frac{1}{n}$. This gives another real solution
that differs starting from the third order
\begin{align}
\Phi_3 =& \frac{1}{72}\Big((X^3,1,1)+9(X^2,X,1)-3(X^2,1,X)-6(X,X^2,1)-6(X,X,X)
\nonumber\\
    &\qquad -3(X,1,X^2)-2(1,X^3,1)-6(1,X^2,X)+9(1,X,X^2)+(1,1,X^3) \Big)\,.
\end{align}
The first choice seems more natural since it does not involve
the scheme changing state $(X,X,X)$.

Yet another possible integration scheme is to integrate
each term of~(\ref{inductionSco}) at the $\mbox{$\Omega$-site}$.
This is not given by a choice of $\vec \al_n$'s, but it is easy to see
that it also leads to a symmetric integration prescription and
therefore to a real solution. The recursion relation can be
written explicitly in this case as
\begin{equation}
\begin{aligned}
\label{recursion}
\Phi_k&=\sum_{n=1}^k \frac{(-1)^n}{2n!}\Big(
   (\partial_{x_0}^{n-1}\Phi_{k-1},X^n)-(X^n,\partial_{x_0}^{n-1}\Phi_{k-1})\Big)\\&+
\sum_{n=1}^k\sum_{l=1}^n\sum_{j=1}^{k-2} \frac{(-1)^n}{4(l-1)!(n-l)!n}
    (\partial_{x_0}^{l-1}\Phi_j,X^n,\partial_{x_0}^{n-l}\Phi_{k-1-j})\,.
\end{aligned}
\end{equation}
The third order term in this case is
\begin{equation}
\Phi_3=\frac{1}{24} \left(3(1,X,X^2)+3(X^2,X,1)-2(1,X^3,1)-6(X,X,X)\right),
\end{equation}
while the fourth order term is
\begin{align}
\nonumber
\Phi_4 & =\frac{1}{96}\big((X^4,1,1,1)-(1,1,1,X^4)\big) +
  \frac{1}{24}\big((1,X,1,X^3)+(1,X^3,X,1)-(X^3,1,X,1) \\ \nonumber & -(1,X,X^3,1)\big)
  +\frac{1}{32}\big((1,1,X^2,X^2)+(X^2,1,X^2,1)-(X^2,X^2,1,1)-(1,X^2,1,X^2)\big)
  \\ & +\frac{1}{16}\big((X,X,X^2,1)+(X,X^2,1,X)+(X^2,1,X,X)
  \\ \nonumber & \quad\ -(1,X^2,X,X)-(X,1,X^2,X)-(X,X,1,X^2)\big)\,.
\end{align}
The closed form expression for the recursion relations~(\ref{recursion})
allows us to calculate
higher order terms. The number of terms in $\Phi_n$
seems to grow exponentially fast.
For $n=1..9$ there are $(1, 2, 4, 16, 43, 152, 521, 1812, 6521)$
summands respectively.

%===========================================================
\section{A map between bosonic and supersymmetric solutions}
%===========================================================
\label{sec:Map}

In this section we show how an $x_0$-independent solution for the
superstring can be built from an $x_0$-independent solution of
the bosonic string, such that~(\ref{PsiG}) holds.
To that end it is useful to define
\begin{equation}
\label{LogGamma}
\Xi_\Gamma(x_0) \equiv (\partial_{x_0}\Gamma) \Gamma^{-1}\,.
\end{equation}
It is interesting to observe that $\Gamma$ is the path-ordered exponential
of $\Xi_\Gamma$
\begin{equation}
\label{ExpXi}
\Gamma = {\cal P}\exp\int^{x_0}\Xi_\Gamma(x) dx
 \equiv 1+\sum_{n=1}^\infty
  \int^{x_0}\Xi_\Gamma(x_1)dx_1\int^{x_1}
   \Xi_\Gamma(x_2)dx_2 \cdots \int^{x_{n-1}}\Xi_\Gamma(x_n)dx_n\,.
\end{equation}
The freedom in defining the above integration goes beyond setting lower
limits to the integrals, as $x_0$ can be related to $X(z)$ insertions
at any point on the boundary, giving a continuum of degrees of freedom.
Restricting the resulting expressions to the form of our ansatz, leaves
us with the same expressions for $\Gamma$ and the same ambiguity
of defining the integration scheme discussed in the previous section.
A similar construction was used in~\cite{Okawa:2007it} to generate a real
solutions from a real $\Xi$.
The difference is that in~\cite{Okawa:2007it}, one
integrates over the gauge parameter $\lambda$ rather than over $x_0$.

The condition for $x_0$-independence
of $\Psi$ is equivalent to the condition that $\Xi_\Gamma$ is
$\mbox{$Q$-closed}$,
since the definition of $\Psi$~(\ref{PsiForm}), implies
\begin{equation}
\label{bosCond}
Q\Xi_\Gamma=\Gamma(\partial_{x_0}\Psi)\Gamma^{-1}=0\,.
\end{equation}
For a supersymmetric solution of the form (\ref{GForm}),
$x_0$-independence implies
\begin{equation}
\label{SUSYcond}
\Xi_\Gamma-\Xi_{\tilde\Gamma}=\tilde\Gamma(\partial_{x_0} G)\Gamma^{-1}=0\,.
\end{equation}
Thus, the condition we were after can be written as
\begin{equation}
\partial_{x_0}\Gamma \Gamma^{-1}=
   \partial_{x_0}\tilde\Gamma \tilde\Gamma^{-1}=\Xi \,,
\end{equation}
where $\Xi$ is arbitrary.

Now, since $\tilde\Phi$ is exact
\begin{equation}
Q\tilde\Phi = 0 \quad\Rightarrow\quad
Q\tilde\Gamma = 0 \quad\Rightarrow\quad
Q\Xi_{\tilde\Gamma}=Q\Xi_\Gamma=0\,.
\end{equation}
Thus, the $x_0$-independence of $G$ results in a closed $\Xi_\Gamma$.
Meaning that the
bosonic condition~(\ref{bosCond}) follows from the supersymmetric
condition~(\ref{SUSYcond}), in accordance with~(\ref{OtherDir}).

The solutions representing the
photon marginal deformation in the bosonic theory in the left, right and
symmetric schemes, all result in the expression
\begin{equation}
\Xi_\Gamma= \lambda\Omega\,.
\end{equation}
This relation is not modified if we replace $\Phi$ by $\tilde\Phi$.
Therefore, all schemes can be used interchangeably
to create supersymmetric solutions.

%=============================================
\section{Supersymmetric marginal deformations}
%=============================================
\label{sec:Susy}

It is not a priori clear to which superstring state the bosonic photon
marginal deformation would be mapped.
First, in~\ref{sec:FirstOrder} we show that the photon state of the
supersymmetric theory can be written as a pure gauge state.
This proves that the full superstring photon marginal deformation
can be generated using our methods.
Then, we demonstrate how to get explicit solutions to all orders
in~\ref{sec:AllOrders}
and real solutions in~\ref{sec:Real}.

%===============================
\subsection{The linear solution}
%===============================
\label{sec:FirstOrder}

Expanding the superstring field
\begin{equation}
G = 1 + \lambda G_1 + \cO(\lambda^2)\,,
\end{equation}
yields the linear order of
the superstring field equation of motion~(\ref{SusyEom})
\begin{equation}
\label{SusyLinEq}
\eta_0 Q G_1 = 0\,.
\end{equation}
The photon state
\begin{equation}
G_1 = c\xi e^{-\phi}\psi(0)\ket{0},
\end{equation}
solves this equation.
Here, $\psi$ has an implicit $\mu$ index and is of conformal weight
$\frac{1}{2}$.
Like in the bosonic case, we would like to write this state as a pure gauge
state generated by a singular gauge transformation. This will allow us
to generate the higher order terms for this solution.
In superstring field theory there are two gauge fields from which
pure gauge states can be built.
Expanding the infinitesimal gauge transformation~(\ref{InfiGauge})
to linear order in $\la$ gives
\begin{equation}
G_1 = -Q\tilde\Lambda_1 + \eta_0\Lambda_1\,.
\end{equation}
Notice that $G_1$ has the $\xi_0$ operator in it, implying that it lies
in the large Hilbert space.

Like in the bosonic case we have to enlarge the Hilbert
space using $x_0$. The $\psi$ operator will be generated thanks to
the $\gamma G_m$ factor in the BRST charge\footnote{Note that we use
the conventions of~\cite{Fuchs:2007yy} for $\partial X$ on the boundary
(eq.~2.5 there), where $\partial$ denotes derivation with respect to the
boundary coordinate $\frac{z+\bar z}{2}$. This
convention results in simple expressions, so we continue to follow it.
This is what we mean by $\partial X$ everywhere, except in the definition
of $G_m$, where we write explicitly $\partial_z X$.
$\partial_z X$ and $\partial X$ differ by a factor of $2$. For the other
operators there is no such issue, because they are holomorphic.}
\begin{equation}
G_m = i\sqrt{2}\psi\partial_z X \quad\Rightarrow\quad
[Q,X]=c\partial X -i \sqrt{2} \eta e^\phi\psi\,.
\end{equation}
Then it is natural to guess
\begin{equation}
\label{GaugeParTil}
\tilde\Lambda_1 = PX(0)\ket{0} \quad\Rightarrow\quad
    \tilde\Phi_1 = Q\tilde\Lambda_1 = X(0)\ket{0} - PQX(0)\ket{0}.
\end{equation}
The first term is redundant and can be canceled by the other gauge field
\begin{equation}
\label{GaugePar}
\Lambda_1 = \xi X(0)\ket{0} \quad\Rightarrow\quad
    \Phi_1 = \eta_0\Lambda_1 = X(0)\ket{0}.
\end{equation}
In total we get
\begin{equation}
G_1 = -\tilde\Phi_1 + \Phi_1 =
PQX(0)\ket{0} = P(c\partial X-i\sqrt{2}\eta e^\phi\psi)(0)\ket{0}
    = c\xi e^{-\phi}\psi(0)\ket{0},
\end{equation}
which is exactly what we want.

%===============================
\subsection{Higher order terms}
%===============================
\label{sec:AllOrders}

To get a solution to the non-linear equation of motion we need
to use the integrated gauge transformation~(\ref{GForm}).
Plugging the first order gauge
parameters~(\ref{GaugeParTil}), (\ref{GaugePar})
into~(\ref{GForm})
produces $x_0$-dependence at higher orders,
no matter what functions $\Gamma(\Phi),\tilde\Gamma(\tilde\Phi)$ are used.
We therefore need to add counter terms.

One could try to use the bosonic $\Phi$, where the counter terms are
known (for example the left scheme solution~(\ref{leftSchemeSol}))
together with the $\tilde \Phi$ defined by~(\ref{Map}).
This gives an $\mbox{$x_0$-dependent}$ solution,
as can be seen by a direct calculation.
Alternatively, we can find and solve a set of
differential equations analogous to the ones of the bosonic case. This
is presented below.

We choose to work in the left scheme, for which we have a closed form
solution in the bosonic theory. Relying on the relation
between the bosonic and supersymmetric solutions
we write $G_L$ in a form similar to $\Psi_L$
\begin{equation}
G_L = \Gamma_L(\tilde\Phi)^{-1}\Gamma_L(\Phi) =
    (1-\tilde\Phi)\frac{1}{1-\Phi}\,.
\end{equation}
In accordance with section~\ref{sec:Map}, we require
that both $\Phi$ and $\tilde\Phi$ satisfy
equations similar to the bosonic case
\begin{equation}
\label{SuperLeft}
\partial_{x_0}\Phi=\lambda(1 -\Phi)\Omega\,,\quad
\partial_{x_0}\tilde\Phi=\lambda(1 -\tilde\Phi)\Omega\,.
\end{equation}
Not surprisingly, this gives an $x_0$-independent solution
\begin{equation}
\partial_{x_0}G_L = 0\,.
\end{equation}
It is insufficient to solve~(\ref{SuperLeft}), since these equations do
not by themselves imply the equation of motion~(\ref{SusyEom}).
We want to find gauge fields $\Lambda,\tilde \Lambda$
generating these $\Phi,\tilde \Phi$ and a solution as in~(\ref{GForm}).
We can write the following equations for the gauge fields
\begin{equation}
\label{Lambda_eq}
\partial_{x_0}\Lambda=\lambda\big(\xi(0)\Omega -\Lambda\Omega\big)\,,\qquad
\partial_{x_0}\tilde\Lambda=\lambda\big(P(0)\Omega -
    \tilde\Lambda\Omega\big)\,,
\end{equation}
from which~(\ref{SuperLeft}) directly follow.
The position of the $P$ and $\xi$ operators was explicitly shown to
emphasize that they operate on the vacuum state and not star-multiply it.
The solution for the gauge fields is
\begin{align}
\label{LambdaLeft}
\Lambda &= \sum_{n=1}^\infty\lambda^n\Lambda_n\,,\qquad
\Lambda_n = -\frac{(-1)^n}{n!}(\xi X^n,\ldots,1)\,,\\
\tilde\Lambda &= \sum_{n=1}^\infty\lambda^n\tilde\Lambda_n\,,\qquad
\tilde\Lambda_n = -\frac{(-1)^n}{n!}(PX^n,\ldots,1)\,,
\end{align}
which yields the string fields
\begin{align}
\Phi &= \eta_0\Lambda = -\sum_{n=1}^\infty\frac{(-\la)^n}{n!} (X^n,\ldots,1)\,, \\
\tilde\Phi &= Q\tilde\Lambda = -\sum_{n=1}^\infty\frac{(-\la)^n}{n!}( X^n
    - n Y X^{n-1} - n(n-1) Z X^{n-2},\ldots,1 )\,.
\end{align}
For this calculation we have used the commutation relation
\begin{equation}
\label{QXn}
[Q,X^n]=-i\sqrt{2}n\eta e^{\phi}\psi X^{n-1} +n c\partial X X^{n-1} -
    n(n-1)\partial c X^{n-2}\,,
\end{equation}
and defined
\begin{equation}
Y \equiv -i\sqrt{2}P\eta e^{\phi}\psi = -i\sqrt{2}c\xi e^{-\phi}\psi\,,
\qquad  Z \equiv c\partial c \xi\partial\xi e^{-2\phi}\,.
\end{equation}
The operator $Z$ has the unique property that all its quantum numbers
are zero.

The field $\Phi$ of the superstring
looks exactly the same as the field $\Phi$ of the bosonic string.
The fact that  $\Phi$ is $\eta_0$ closed
means that the related gauge transformation can be simply written
as in~(\ref{SUSYbos}),
\begin{equation}
\Lambda=\xi_0\Phi\,.
\end{equation}
This state does not obey~(\ref{Lambda_eq}), but it only
differs from~(\ref{LambdaLeft}) by an $\eta_0$-closed term.
Thus, both gauge fields result in exactly the same solution.

%==========================
\subsection{Real solutions}
%==========================
\label{sec:Real}

We now want to identify a real solution.
The reality condition for the superstring field is
\begin{equation}
G^* = G^{-1}\,.
\end{equation}
The fields $X$, $Y$ and $Z$ are all real.
We assume that $\Phi$ and $\tilde\Phi$ are chosen such
that they keep the imaginary nature of their lowest order.
Then for $G$ to be real, the functions that generate it need to
satisfy~(\ref{RealCond}) just as in the bosonic case.
The next step is to imitate the bosonic symmetric solution
\begin{equation}
G = \Gamma_S(\tilde\Phi)^{-1}\Gamma_S(\Phi) =
    \frac{1-\frac{1}{2}\tilde\Phi}{1+\frac{1}{2}\tilde\Phi}\;
    \frac{1+\frac{1}{2}\Phi}{1-\frac{1}{2}\Phi}\,,
\end{equation}
and require
\begin{equation}
\partial_{x_0}\Phi = \lambda(1-\frac{1}{2}\Phi)\Omega(1+\frac{1}{2}\Phi)\,,
\qquad  \partial_{x_0}\tilde\Phi =
    \lambda(1-\frac{1}{2}\tilde\Phi)\Omega(1+\frac{1}{2}\tilde\Phi)\,,
\end{equation}
to get an $x_0$-independent solution.
Just like in the left scheme solution, the expression for the supersymmetric
$\Phi$ is the same as that of $\Phi$ of the bosonic string.
For $\tilde\Phi$ we need to solve the equation
\begin{equation}
\label{SUSYeq}
\partial_{x_0}\tilde\Lambda = \lambda\Big( P(0)\Omega -
    \frac{1}{2}\tilde\Lambda\Omega + \frac{1}{2}\Omega\tilde\Lambda -
    \frac{1}{8}\tilde\Lambda\Omega(Q\tilde\Lambda) -
    \frac{1}{8}(Q\tilde\Lambda)\Omega\tilde\Lambda
    \Big)\,,
\end{equation}
where we have chosen the symmetric form of the equation.

We can use an integration choice analogous to that of the bosonic
case~(\ref{recursion}) of integrating at the $\Omega$ site.
This can be explicitly written as
\begin{align}
\nonumber
\tilde\La_k&=\sum_{n=1}^k \frac{(-1)^n}{2n!}\Big(
   (\partial_{x_0}^{n-1}\tilde\La_{k-1},X^n)
    -(X^n,\partial_{x_0}^{n-1}\tilde\La_{k-1})\Big)+
\sum_{n=1}^k\sum_{l=1}^n\sum_{j=1}^{k-2} \frac{(-1)^n}{8(l-1)!(n-l)!n}\cdot\\&\cdot
\label{SUSYrecursion}
    \Big((\partial_{x_0}^{l-1}Q\tilde\La_j,X^n,\partial_{x_0}^{n-l}\tilde\La_{k-1-j})+
    (\partial_{x_0}^{l-1}\tilde\La_j,X^n,\partial_{x_0}^{n-l}Q\tilde\La_{k-1-j})\Big)\,.
\end{align}
This results in
\begin{align}
\tilde\La_2=\frac{1}{4}\Big((P,X^2)-(X^2,P)\Big)+\frac{1}{2}\Big((X,PX)-(PX,X)\Big)\,.
\end{align}
Note that unlike for the bosonic case, there is a freedom in choosing
a real solution already at the second order since the location of the $P$ insertion
should be specified.

It is possible to simplify the expressions by choosing a different
integration prescription, namely to integrate in the location of the $P$ insertion.
It should be understood that the $P$'s appearing in
expressions that result from $Q\tilde\La$ in~(\ref{SUSYeq}) are not the ones
where integration should be performed, since a $Q$ is acting on them.
With this understanding, every $\Phi_k$ has exactly one site with
a $P$ insertion and our algorithm is well-defined.
The second order result is then
\begin{align}
\tilde\Lambda_2 &=\frac{1}{4}\Big((1,PX^2)-(PX^2,1)\Big)\,,\\
\tilde\Phi_2 &= \frac{1}{4}\Big((1,X^2-2XY-2Z)-(X^2-2XY-2Z,1)\Big)\,,\\
G_2 &= \frac{1}{2}\Big((1,XY)-(XY,1)+(Y,X)-(X,Y)+(Y,Y)
    -(Z,1)+(1,Z)\Big)\,,
\end{align}
where for the evaluation of $G_2$ we have taken the expression for $\Phi_2$
from the bosonic string.
Calculating higher order terms is straightforward, but not very illuminating.

%===========================================
\section{The universal superstring solution}
%===========================================
\label{sec:Universal}

Schnabl's original solution for the bosonic string~\cite{Schnabl:2005gv}
can also be written as a gauge transformation~\cite{Okawa:2006vm}
\begin{equation}
\Psi_\lambda = (1-\Phi)Q\frac{1}{1-\Phi}\,,\qquad
\Phi = \frac{\la}{\pi} \BB_0^\dag c(0)\ket{0}.
\end{equation}
$\Phi$ may also be viewed as a singular gauge transformation since
it generates a field which is both exact and satisfies the Schnabl gauge
\begin{equation}
\BB_0 Q\Phi = 0\,.
\end{equation}
The Siegel gauge does not seem to permit such states.

Let us define a similarity transformation like
the one we used for regularizing the three-vertex~\cite{Fuchs:2006gs}
\begin{equation}
\BB_0^s\equiv s^{-L_0}\BB_0 s^{L_0}\,, \qquad
\Phi_s \equiv s^{-L_0}\Phi\,.
\end{equation}
For any finite $s$, states in the Schnabl gauge transform into states
in the $\BB_0^s$ gauge. In the limit $s\rightarrow0$ we reach the
Siegel gauge. All physical states transform from the Schnabl gauge
to the Siegel gauge, but the state $\Phi$ is singular in this limit,
due to the singularity in this limit of $\BB_0^\dag$ (and $\cL_0^\dag$).

A conceptual difference between Schnabl's universal solution and
our marginal deformation is that for small $\lambda$ his solution
is indeed a pure gauge solution. Only at the critical value
$\lambda=1$ does it become a physical solution.
Still, we can speculate that the relation between bosonic and
superstring solutions also holds for this case.
The state
\begin{equation}
G_\lambda = (1-\tilde\Phi)\frac{1}{1-\Phi}\,,\qquad
\Phi = \frac{\lambda}{\pi} \BB_0^{\dagger}c(0)\ket{0},\quad
\tilde\Phi = QP(0)\Phi\,.
\end{equation}
is clearly a solution to the superstring field equation of motion.
$\Phi$ was copied from the bosonic string and since it is built
upon the vacuum state, there seems to be no ambiguity about the
location of the $P$ insertion in $\tilde\Phi$.
One can check that $G_\lambda$ satisfies Schnabl's gauge
\begin{equation}
\BB_0 (G_\lambda - 1) = 0\,.
\end{equation}

We suggest that at the critical value of $\lambda$
this could be the universal solution for the superstring.
Generically, there is no tachyon in superstring field theory, so we should
not think of this state as being the tachyon vacuum.
We believe that this solution represents a state with no $D$-branes
and therefore has an empty cohomology.

Like in the bosonic case, the study of this state should require some
kind of regularization. We leave this study for future work.

%====================
\section{Conclusions}
%====================
\label{sec:Conclusions}

It seems that all known solutions to bosonic and supersymmetric string
field theories can be written as pure gauge solutions.
The difference between different solutions and different approaches is
in the choice of the gauge field.
The approach of this paper and~\cite{Fuchs:2007yy} gives elegant results
that generalize automatically to singular currents, but
works only for the photon operator.
The approach
of~\cite{Kiermaier:2007ba,Schnabl:2007az,Erler:2007rh,Okawa:2007ri,Okawa:2007it}
works for all non-singular currents, but requires complicated counter
terms for handling singular currents.
The generalization of our approach to other operators
was discussed in~\cite{Fuchs:2007yy}.
It would be interesting to complete this program.

%=========================
\section*{Acknowledgments}
%=========================

We would like to thank Sudarshan Ananth, Rob Potting and Stefan Theisen
for useful discussions.
The work of M.~K.\ is supported by a Minerva fellowship.
The work of E.~F.\ is supported by the German-Israeli Project
cooperation (DIP H.52).

\appendix

%===============================
\section{Split string formalism}
%===============================

In order to compare our solution to that obtained by other authors
it may be useful to write it using the formalism of~\cite{Erler:2007rh}.
Insertion of $X^n$ will be described by an
insertion over the identity string field,
\begin{align}
X^n\equiv X^n\ket{1}=\underbrace{X\star\ldots\star X}_{n\text{ times}}.
\end{align}
Normal ordering in this expression is implicit,
therefore the r.h.s. cannot be strictly viewed as a chain of
matrix multiplications.

Then, between
any two insertion sites there is a strip of string that can be
represented by $F^2=\Omega$.
For example, the bosonic left solution is given by
\begin{equation}
1-\Phi=1+F \sum_{k=1}^\infty \frac{(-\la)^k}{k!} X^k F^{2k-1}\equiv
F \sum_{k=0}^\infty \frac{(-\la)^k}{k!} X^k F^{2k-1}\,.
\end{equation}
This can be written in short as
\begin{equation}
1-\Phi=F e^{\partial_\al\partial_\beta}
   e^{-\al \la X}e^{\beta \Omega}\Big|_{\al=\beta=0}F^{-1}\,.
\end{equation}
Using the bosonic part of~(\ref{QXn}) we can write the solution as
\begin{equation}
\Psi=\la F c\partial X F^{-1} (1-\Phi) \Omega (1-\Phi)^{-1}+
   \la^2 F \partial c F^{-1} (1-\Phi) \Omega^2 (1-\Phi)^{-1}\,.
\end{equation}

%===================================
\section{Integrated strip formalism}
%===================================

The marginal deformations
in~\cite{Kiermaier:2007ba,Schnabl:2007az,Okawa:2007ri,Okawa:2007it}
were all based on the fact that the inverse of $\cL_0$ can be written
as an integration over the width of a strip of string.
Here, we demonstrate that these solution can also be viewed as
pure gauge solutions.

For the bosonic string, our solution is based on the fact that $\Psi_1$,
which is closed
\begin{equation}
Q\Psi_1=0\,,
\end{equation}
can be written as an exact state
\begin{equation}
\Psi_1=Q\Phi_1\,.
\end{equation}
Using the integrated strip one can define the state $J$, which satisfies
\begin{equation}
QJ=1\,.
\end{equation}
This state is defined as an integral of a wedge state with length varying
between zero and $\frac{\pi}{2}$. Since $\frac{\pi}{2}$ is the length of the
local coordinate patch, the integral is over states which remove string strips
and as such, are not generally defined. The expression we get for the physical
state is, however, well defined.
We can use this state to write
\begin{equation}
\Phi_1=J\Psi_1 \quad\Rightarrow\quad Q\Phi_1=(QJ)\Psi_1-J(Q\Psi_1)=\Psi_1\,,
\end{equation}
which is exactly what we need.
It is a formal gauge field, not strictly existing due to the appearance of
$J$ in its definition and its gauge variation gives the correct first order
solution. These are also the properties of our gauge field $x_0\ket{0}$.

The full solution, ignoring the issue of singular OPE's is
\begin{equation}
\Psi_n = (Q\Phi_1)\Phi_1^{n-1} = \Psi_1(J\Psi_1)^{n-1}\,.
\end{equation}
This is exactly the form of the solution of~\cite{Okawa:2007ri}.
The structure of this solution is like ours,
yet the states involved are different. Specifically,
\begin{equation}
X(0)\ket{0} \neq J c\partial X(0)\ket{0},
\end{equation}
and the solutions differ, but are presumably gauge equivalent.

The supersymmetric theory requires a different $J$ state.
This time $J$ satisfies the relation
\begin{equation}
Q\eta_0 J = 1\,.
\end{equation}
We can use this state to write
\begin{align}
\Lambda = \lambda(Q G_1)J \quad&\Rightarrow\quad
    \Phi = \eta_0\Lambda = -\lambda(Q G_1)(\eta_0 J)\,,\\
\tilde\Lambda = -\lambda G_1(\eta_0 J) \quad&\Rightarrow\quad
    \tilde\Phi = Q\tilde\Lambda = -\lambda G_1 - \lambda(Q G_1)(\eta_0 J)\,,
\end{align}
where we used~(\ref{SusyLinEq}).
This means that every state of the form
\begin{equation}
\label{GeneralSolution}
G = \tilde\Gamma(\tilde\Phi)^{-1} \Gamma(\Phi)\,,
\end{equation}
solves the equation of motion, with the right linear term $G_1$,
if the functions $\Gamma,\tilde\Gamma$ are of the form~(\ref{fForm}).
The superstring marginal solution of~\cite{Okawa:2007ri},
\begin{equation}
G^{-1} = 1-\frac{\lambda}{1-\lambda(Q G_1)(\eta_0 J)} G_1\,,
\end{equation}
is reproduced by choosing
\begin{equation}
\Gamma(\Phi) = 1+\Phi\,,\qquad
\tilde\Gamma(\tilde\Phi) = 1+\tilde\Phi\,.
\end{equation}

%================
\bibliography{FK}
%================

\end{document}